\documentclass[pre,twocolumn,showpacs,preprintnumbers,amsmath,amssymb]{revtex4}
\usepackage{bm,graphicx}

\begin{document}

\preprint{Accepted for publication in Phys.\ Rev.\ E}

\title{Interaction and flocculation of spherical colloids\\ wetted by a
surface-induced corona of paranematic order}

\author{Paolo Galatola}
\email{galatola@ccr.jussieu.fr}
\affiliation{LBHP (UMR 7057), Universit\'e Paris 7---Denis Diderot and
FR CNRS 2438 ``Mati\`ere et Syst\`emes Complexes", Case
7056, 2 place Jussieu, F-75251 Paris cedex 05, France}

\author{Jean-Baptiste Fournier}
\email{jbf@turner.pct.espci.fr}
\affiliation{Laboratoire de Physico-Chimie Th\'eorique and
FR CNRS 2438 ``Mati\`ere et Syst\`emes Complexes",
ESPCI, 10 rue Vauquelin, F-75231 Paris cedex 05,
France}
\author{Holger Stark}
\email{holger.stark@uni-konstanz.de}
\affiliation{Universit\"at Konstanz, Fachbereich Physik, D-78457
Konstanz, Germany}

\date{\today}

\begin{abstract}
Particles dispersed in a liquid crystal above the nematic-isotropic
phase transition are wetted by a surface-induced corona of paranematic
order. Such coronas give rise to pronounced two-particle interactions.
In this article, we report details on the analytical and numerical study
of these interactions published recently [Phys.\ Rev.\ Lett.\
\textbf{86}, 3915 (2001)]. We especially demonstrate how for large
particle separations the asymptotic form of a Yukawa potential arises.
We show that the Yukawa potential is a surprisingly good description for
the two-particle interactions down to distances of the order of the
nematic coherence length. Based on this fact, we extend earlier studies
on a temperature induced flocculation transition in electrostatically
stabilized colloidal dispersions [Phys.\ Rev.\ E \textbf{61}, 2831
(2000)].  We employ the Yukawa potential to establish a flocculation
diagram for a much larger range of the electrostatic parameters, namely
the surface charge density and the Debye screening length. As a new
feature, a kinetically stabilized dispersion close to the
nematic-isotropic phase transition is found.
\end{abstract}


\maketitle


\section{\label{intro}Introduction}

Colloidal dispersions are suspensions in a host fluid of solid or liquid
particles, of radius ranging from $10\,\mathrm{nm}$
to~$10\,\mu\mathrm{m}$~\cite{Hunter}. They are long-lived metastable
states of matter that present great interest both from a fundamental
point of view and in applications, as, e.g., in paints, coatings, foods
and drugs. Their stability against flocculation is a key issue in
colloidal physics, since the properties of a colloidal dispersion
drastically change when a transition from a dispersed to an aggregate
state occurs: to prevent coagulation due to attractive van der Waals
forces, colloidal particles are usually treated in order to produce
electrostatic or steric repulsive interactions.

Recently, great attention was paid to liquid crystal colloidal
suspensions and emulsions, i.e., dispersions of solid particles or
liquid droplets---respectively---in a liquid
crystal~\cite{Terentjev1995,poulin97a,poulin97b,poulin98,Stark2001}.  In
the case of emulsions in a nematic phase, the radial anchoring of the
nematic molecules at the surface of the droplets yields topological
defects of the nematic texture in the vicinity of the droplets. These
defects produce strong repulsions that stabilize the liquid droplets
against coalescence~\cite{poulin97b}.  Similar effects are present also
in other anisotropic fluid hosts, such as lyotropic solutions of
anisotropic micelles~\cite{poulin99}, and in different liquid crystal
phases, as in cholesterics~\cite{zapotocky99}. Such systems form
composite materials with unusual
properties~\cite{Meeker2000,anderson01,loudet00}. 

Even more recently, there has been a growing interest in the
pretransitional surface-induced interactions mediated by the paranematic
order arising in the vicinity of surfaces at temperatures above the
nematic--isotropic phase transition. In~\cite{borstnik97}, the force
between two parallel plates immersed in a liquid crystal slightly above
the bulk isotropic--nematic transition was theoretically investigated.
In~\cite{kocevar01}, the force between a flat surface and a glass
microsphere, mediated by the surface-induced paranematic order, was
experimentally measured using an atomic force microscope. The
measurements were interpreted in terms of the force between two flat
surfaces using the Derjaguin approximation~\cite{Israelachvili92}.

Interesting effects regarding the stabilization of colloids are thus to
be expected in the \textit{isotropic} phase of a nematogenic material
when the colloids are wetted by a corona of paranematic
order~\cite{borstnik99,galatola01,Stark2002}. Two effects compete: an
attraction due to the favorable overlapping of the paranematic halos
(which reduces the volume of the thermodynamically unvaforable
paranematic phase) and a repulsion due to the distortion of the director
field. The vicinity of a phase transition may give a critical character
to the stabilization mechanism and yield rich reversible
phase-separation behaviors, as predicted in~\cite{lowen95,Bauer2000} for
a simpler system with a scalar order-parameter.  In
Ref.~\cite{borstnik99}, the interaction of two spherical particles
immersed in a liquid crystal above the nematic-isotropic phase
transition was investigated using a uniaxial ansatz function for the
paranematic order, in the limit of sphere radii large with respect to
the nematic coherence length. This approximate solution was used in
Ref.~\cite{borstnik2000} to analyze the stability of a suspension of
such colloidal particles, in the presence of destabilizing van der Waals
attractive interactions and stabilizing electrostatic repulsions.  In
Ref.~\cite{galatola01}, an exact numerical solution for the above
paranematic interaction was obtained, using a multipolar expansion of
the tensor order-parameter and taking into account biaxiality. The
possibility of stabilizing colloidal particles was discussed,
emphasizing the case of rather small particles, i.e., of size comparable
to several nematic coherence lengths.  A large-separation asymptotic
analytic form of Yukawa type was also obtained, that was shown to very
closely describe the exact interaction between large particles up to
separations comparable to the nematic coherence
length~\cite{fournier02}.  Alternative derivations of the Yukawa
potential were given in Ref.~\cite{Stark2002} based on a Debye-H\"uckel
type approximation and a geometric view.

In this work, we revise in detail the solution found in
Ref.~\cite{galatola01}, which is exact in the limit of weak
surface-induced paranematic order. We give a particular emphasis on the
meaning and limits of validity of the approximate asymptotic solution.
The latter is used to analyze in detail the stability of a colloidal
suspension in the presence of van der Waals and electrostatic
interactions. We obtain and discuss general diagrams for the stability
of the suspension as a function of the various relevant parameters.

The detailed plan of our paper is the following. In Section~\ref{model}
we describe our Landau-de Gennes~\cite{deGennes_book} model for the bulk
and the surface free-energy. In Section~\ref{equilibrium_equations} we
present the corresponding equilibrium equations and formulate the
general multipolar expansion for the bulk order-parameter in spherical
coordinates. This expansion is used in Section~\ref{sec:one} to obtain
the exact solution for an isolated spherical particle. The interaction
of two such particles is discussed in Section~\ref{sec:two}. In
particular, in Section~\ref{sec:asympt} we derive the asymptotic
interaction energy. The asymptotic solution is analyzed in
Section~\ref{sec:yuk} in terms of the superposition of single-particle
solutions. In Section~\ref{sec:derj} we discuss the limits of validity
of the Derjaguin approximation. In Section~\ref{subsec.inter.num} we
present an exact numerical solution for the interaction of two
particles.  The comparison between the asymptotic interaction energy and
the exact numerical one is performed in Section~\ref{inter.num.ener}. In
Section~\ref{sec:texture} we discuss the texture between two interacting
particles, and in Section~\ref{sec:defects} we analyze the defect ring
that appears in the paranematic texture between the two particles.
Finally, in Section~\ref{sec:floc}, using our asymptotic solution, we
discuss the stability of such colloidal dispersions.

\section{Model} \label{model}

Let us first describe the nematogenic phase in which the colloids will
be placed. Nematic liquid crystals are anisotropic liquid phases, in
which elongated molecules display a long-range orientational order. This
order is described by a symmetric traceless tensorial
order-parameter~$Q_{ij}$ ($i,j=1,2,3$), since nematics are
non-polar~\cite{deGennes_book}. The eigenvectors of~$Q_{ij}$ represent
the axes of main molecular orientation and its eigenvalues describe the
amount of orientational ordering in each direction.  Usually, nematics
are uniaxial phases, however biaxiality naturally arises in
inhomogeneous situations, e.g., in the vicinity of
defects~\cite{kralj99}. The Landau-de Gennes expansion of the bulk
free-energy density~\cite{deGennes_book} has the
form~\cite{vertogen_book,kralj92} 
\begin{eqnarray}
\label{fb}
f_b=&&
\frac{a}{2}\,Q_{ij}Q_{ij}-\frac{b}{3}\,Q_{ij}Q_{jk}Q_{ki}
+\frac{c}{4}\,\left(Q_{ij}Q_{ij}\right)^2\nonumber\\
&&+\frac{L}{2}\,Q_{ij,k}Q_{ij,k}.
\end{eqnarray}
Summation over repeated indices is implied and comma indicates
derivation. The coefficients are such that $a=\alpha(T-T^\ast)$, with
$\alpha$, $c$, and~$L$ positive. The temperature $T^\ast$ is the lowest
temperature $T$ at which the isotropic phase can exist. The presence of
the cubic term $Q_{ij}Q_{jk}Q_{kl}$ in the homogeneous part implies that
the nematic--isotropic transition is first-order.  To simplify the
description, we have introduced only one gradient term in
Eq.~(\ref{fb}), which corresponds to the usual one-constant
approximation~\cite{deGennes71}.  The most general expansion of the
surface free-energy density is~\cite{gossens85,nobili92} 
\begin{eqnarray}
f_s=&&v_1Q_{ij}\nu_i\nu_j+\frac{v_{21}}{2}\,Q_{ij}Q_{ij}
+\frac{v_{22}}{2}\,Q_{ij}Q_{jk}\nu_i\nu_k\nonumber\\*
&&+\,\frac{v_{23}}{2}\,Q_{ij}Q_{kl}\nu_i\nu_j\nu_k\nu_l\,
+\mathcal{O}(Q^3),
\end{eqnarray}
where the $\nu_i$'s ($i=1,2,3$) are the components of the outward
normal~$\bm{\nu}$ of the surface.  Owing to the linear term, the surface
always locally favor a nonzero order.

Since we are dealing with the isotropic phase, and, in addition, we
assume a weak surface-induced order and temperatures not too close to
the phase transition, we can neglect the third and higher-order terms in
$f_b$ and $f_s$~\cite{borstnik99,galatola01}. Then, the free-energy
being quadratic, exact calculations are feasible. To simplify, we shall
also retain only the simplest quadratic surface term, by setting
$v_{22}=0$ and $v_{23}=0$. Then the free-energy densities can be put in
the form
\begin{subequations}
\label{eq:qfe}
\begin{eqnarray}
f_b&=&\frac{a}{2}Q_{ij}Q_{ij}+\frac{L}{2}Q_{ij,k}Q_{ij,k},\\
f_s&=&\frac{1}{2} W \left( Q_{ij} - Q_{ij}^{(0)}\right)
\left( Q_{ij} - Q_{ij}^{(0)}\right),
\label{eq:fs}
\end{eqnarray}
\end{subequations}
where 
\begin{equation}
\label{Q0}
Q^{(0)}_{ij}=S_0\left(\nu_i\nu_j-\frac{1}{3}\delta_{ij}\right)
\end{equation}
is the preferred order-parameter at the surface.  The surface
free-energy density~(\ref{eq:fs}) is compatible with the experimentally
measured anchoring in the nematic phase~\cite{nobili92}.

In the following, unless otherwise specified, we shall normalize lengths
with respect to the nematic correlation length $\xi=(L/a)^{1/2}$ and
energies with respect to $F_0=a\xi^3 S_0^2$. Because of the conditions
that $Q_{ij}$ must be symmetric and traceless, the free-energy
densities~(\ref{eq:qfe}) are not diagonal in the five independent
components of $Q_{ij}$. Defining the following set of components:
\begin{subequations}
\label{q-components}
\begin{eqnarray}
q_0 &=& \frac{Q_{xx} - Q_{yy}}{S_0},\\
q_1 &=& \frac{Q_{yz}}{S_0}=\frac{Q_{zy}}{S_0},\\
q_2 &=& \frac{Q_{xz}}{S_0}=\frac{Q_{zx}}{S_0},\\
q_3 &=& \frac{Q_{xy}}{S_0}=\frac{Q_{yx}}{S_0},\\
q_4 &=& \frac{Q_{zz}}{S_0},
\end{eqnarray}
\end{subequations}
the normalized free-energy takes the following \textit{diagonal} form:
\begin{equation}
\label{Fen}
F = \sum_{i=0}^4\! c_i\!\left\{
\int\!\! \left[ q_i^2 + \left(\nabla q_i\right)^2 \right]\!
d^3 r
+ w \!\!\int\!\!\! \left(q_i-q_i^{(0)}\right)^2\! d^2 r
\right\},
\end{equation}
where $c_0 = 1/4$, $c_1 = c_2 = c_3 = 1$, $c_4 = 3/4$, and $w = W/a\xi$
is the normalized anchoring strength. The first integral is over the
bulk and the second one over the surface of each colloidal particle.
Thanks to this transformation, the tensorial problem associated with the
paranematic order between the colloids is reduced to the superposition
of five independent scalar problems. 

\section{\label{equilibrium_equations}Equilibrium equations and
solutions}

At equilibrium, the order-parameter texture minimizes the
free-energy~(\ref{Fen}). By setting to zero the variation $\delta F$ of
the free-energy, associated with arbitrary infinitesimal variations of
the~$q_i$ components, we obtain the equilibrium equations
\begin{equation}
\label{eq:bulk}
(\nabla^2-1) q_i = 0,
\end{equation}
in the bulk, and
\begin{equation}
\label{eq:surface}
\bm{\nu}\cdot\nabla q_i = w\left(q_i - q_i^{(0)}\right),
\end{equation}
on the surface of each colloidal particle. Determining the paranematic
order outside the colloidal particles is therefore somewhat similar to
solving an electrostatic problem with mixed boundary conditions in which
the potential is replaced by the $q_i$ fields and the standard Laplacian
operator by $\nabla^2-1$. We shall therefore use multipolar expansions.
Actually, the operator $\nabla^2-1$ is the operator associated with a
massive boson field, which implies a short-range
interaction~\cite{YukawaPotential}.

\subsection{Multipolar expansion}
\label{sec:multi}

In spherical coordinates $(r,\theta,\phi)$, the spherical harmonics
$Y_{\ell m}(\theta,\phi)$ (for their definition see
Appendix~\ref{sphharm}) are eigenfunctions of the angular part of
the Laplacian operator $\nabla^2$ and form a complete basis. Then, each
of the $q_i$'s can be expanded as
\begin{equation}\label{expansion_multipolaire}
q_i(r,\theta,\phi)=\sum_{\ell=0}^\infty \sum_{m=-\ell}^\ell q^{\ell m}_i
\,
u_\ell(r) Y_{\ell m}(\theta,\phi),
\end{equation} 
where the functions $u_{\ell}(r)$ obey the equations
\begin{equation}\label{equation_u}
\frac{d}{dr}\left(r^2\frac{du_\ell}{dr}\right)=\left[r^2+
\ell(\ell+1)\right]u_\ell,
\end{equation}
which follows by inserting expansion\ (\ref{expansion_multipolaire}) 
into the bulk equation\ (\ref{eq:bulk}).
The solution of Eq.~(\ref{equation_u}) that is regular at infinity can
be expressed as
\begin{equation}
u_\ell(r)=\sqrt{\frac{2}{\pi r}}\,K_{\ell+1/2}(r)
\end{equation}
where the $K_{\ell+1/2}(r)$ are modified Bessel functions of half-integer
order. Explicitly,
\begin{subequations}
\label{ui}
\begin{eqnarray}
u_0(r)&=&\frac{e^{-r}}{r},\\
u_1(r)&=&\frac{e^{-r}}{r}\left(1+\frac{1}{r}\right),\\
\label{u2}
u_2(r)&=&\frac{e^{-r}}{r}\left(1+\frac{3}{r}+\frac{3}{r^2}\right),\\
u_{\ell+1}(r)&=&\frac{2\ell+1}{r}u_\ell(r)+u_{\ell-1}(r)
\text{ for }\ell\ge1.
\end{eqnarray}
\end{subequations}
With such an expansion, the paranematic order is thus completely
described by the set of coefficients~$q^{\ell m}_i$.

\subsection{Equilibrium free-energy}

By integrating by parts Eq.~(\ref{Fen}) and using the equilibrium
equations (\ref{eq:bulk}) and~(\ref{eq:surface}), the normalized
free-energy~(\ref{Fen}) associated with an equilibrium solution can be
written as
\begin{equation}
\label{eq:fen}
F = w \sum_{i=0}^4 c_i \int q_i^{(0)} \left(
q_i^{(0)} -q_i\right)d^2 r,
\end{equation}
where the integrations run over the surfaces of all the particles. 

\section{\label{one_particule}Paranematic order around one particle}
\label{sec:one}

Let us consider a spherical particle of reduced radius~$R$. In spherical
coordinates $(r,\theta,\phi)$, the normal to the surface of the particle
is given by $\bm{\nu}=\sin\theta\cos\phi\,\mathbf{\hat x}+
\sin\theta\sin\phi\,\mathbf{\hat y}+\cos\theta\,\mathbf{\hat z}$.
Therefore, the preferred surface order-parameter~(\ref{Q0})
has the following $q$-components [see Eqs.~(\ref{q-components})]
\begin{subequations}
\label{qi0}
\begin{eqnarray}
q^{(0)}_0&=&\sin^2\theta\cos2\phi=
4\sqrt{\frac{2\pi}{15}}\,Y_{22}^R(\theta,\phi)\\
q^{(0)}_1&=&\frac{1}{2}\sin2\theta\sin\phi=
-2\sqrt{\frac{2\pi}{15}}\,Y_{21}^I(\theta,\phi)\\
q^{(0)}_2&=&\frac{1}{2}\sin2\theta\cos\phi=
-2\sqrt{\frac{2\pi}{15}}\,Y_{21}^R(\theta,\phi)\\
q^{(0)}_3&=&\frac{1}{2}\sin^2\theta\sin2\phi=
2\sqrt{\frac{2\pi}{15}}\,Y_{22}^I(\theta,\phi)\\
q^{(0)}_4&=&\cos^2\theta-\frac{1}{3}=
\frac{4}{3}\sqrt{\frac{\pi}{5}}\,Y_{20}(\theta,\phi),
\end{eqnarray}
\end{subequations}
where
\begin{subequations}
\begin{eqnarray}
Y_{\ell m}^R&=&
\frac{1}{2}\left(Y_{\ell m}+Y^*_{\ell m}\right),\\
Y_{\ell m}^I&=&\frac{1}{2i}\left(Y_{\ell m}-Y^*_{\ell m}\right)
\end{eqnarray}
\end{subequations}
are the real and imaginary part, respectively, of the spherical
harmonics. The boundary conditions~(\ref{eq:surface}) become simply
\begin{equation}
\label{bc_one_particle}
\left.\frac{\partial q_i}{\partial r}\right|_{r=R}=
w\left[q_i(R,\theta,\phi)-q_i^{(0)}(\theta,\phi)\right].
\end{equation}
Replacing in Eq.~(\ref{bc_one_particle}) $q_i$ by its multipolar
expansion~(\ref{expansion_multipolaire}) and $q_i^{(0)}$ by its
expression given in Eqs.~(\ref{qi0}), and identifying the coefficients of
the spherical harmonics, yields the solution
\begin{equation}
\label{q-solution_1particule}
\begin{pmatrix}
q_0\\ q_1\\ q_2\\ q_3\\ q_4
\end{pmatrix}
=\sqrt{\frac{2\pi}{15}}
\begin{pmatrix}
2\,Y_{22}^R(\theta,\phi)\\ 
-Y_{21}^I(\theta,\phi)\\ 
-Y_{21}^R(\theta,\phi)\\ 
Y_{22}^I(\theta,\phi)\\ 
\sqrt{\frac{2}{3}}\,Y_{20}(\theta,\phi)
\end{pmatrix}
2\mathcal{A}(R,w)e^{R}\,u_2(r),
\end{equation}
where the amplitude $\mathcal{A}(R,w)$ is given by
\begin{equation}
\mathcal{A}=\frac{R^4w}{R^3(w+1)+R^2(3w+4)+3R(w+3)+9}
\label{4.10}
\end{equation}
and $u_2(r)$ is given by Eq.~(\ref{u2}).

\begin{figure}
\includegraphics[width=.8\columnwidth]{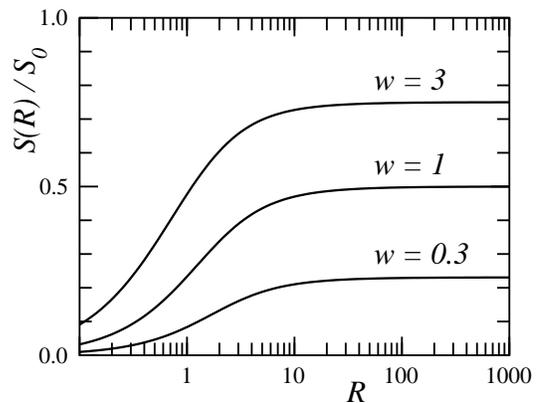}
\caption{\label{s_1bille}Normalized scalar order-parameter $S(R)/S_0$ at
the surface of one isolated spherical colloid as a function of its
normalized radius $R$ for various reduced anchoring strength~$w$.}
\end{figure}

Since Eq.~(\ref{q-solution_1particule}) has the same angular structure
as  Eqs.~(\ref{qi0}), the order-parameter is uniaxial with a radial
director $\mathbf{n}=\bm{\nu}$, i.e., $Q_{ij}=S(r)(\nu_i
\nu_j-\frac{1}{3}\delta_{ij})$, where 
\begin{equation}
S(r)=S_0\mathcal{A}(R,w)
\left(1+\frac{3}{r}+\frac{3}{r^2}\right)
\frac{e^{R-r}}{r}
\label{single}
\end{equation}
is the uniaxial scalar order-parameter. Note that whatever the size of
the particle, the paranematic order relaxes on the coherence
length~$\xi$ (equal to unity in reduced units).  The behavior of the
scalar order-parameter at the surface of the particle ($r=R$), as a
function of the radius~$R$, is shown in Fig.~\ref{s_1bille}. For large
values of $R$, the surface order tends to $S(R)\simeq S_0w/(1+w)$ which
corresponds to the limit of a planar interface. For $R\alt10$ (in units
of $\xi$), the surface order-parameter is reduced because of the energy
cost associated with the splay of the director.

The free-energy of the particle is obtained from Eq.~(\ref{eq:fen}).
With the help of Eqs.~(\ref{qi0}) and~(\ref{q-solution_1particule}) and the
orthonormality of the spherical harmonics, we find 
\begin{equation}
\label{eq:f1}
F_1 = 
\frac{4\pi}{3} \, \mathcal{A}(R,w) \, \left(
R + 4 + \frac{9}{R} + \frac{9}{R^{2}}
\right).
\end{equation}

\section{Interaction of two particles}
\label{sec:two}

\begin{figure}
\includegraphics[width=.45\columnwidth]{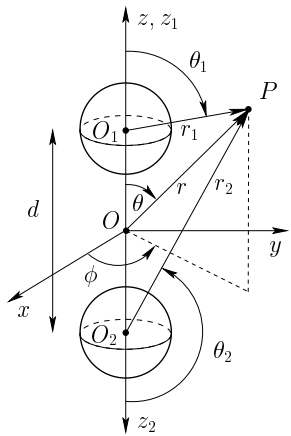}
\caption{\label{2billes}Geometry for the calculation of the paranematic
interaction between two identical spherical colloidal particles
separated by a center-to-center distance~$d$. The origin of the $z$-axis
is in the midpoint~$O$ between the centers of the two particles, that of
$z_1$ in the center~$O_1$ of the upper particle, and that of~$z_2$ in
the center~$O_2$ of the lower one.}
\end{figure}

Let us now consider two identical spherical particles of reduced
radius~$R$ separated by a center-to-center distance~$d$. We introduce
three spherical coordinate systems: a global one, $(r,\theta,\phi)$,
symmetrically placed with respect to the particles, and two local ones,
$(r_1,\theta_1,\phi)$ and $(r_2,\theta_2,\phi)$, centered on the two
particles, as indicated in Fig.~\ref{2billes}. 

\subsection {General formalism}

Using the symmetry of the system allows to reduce the number of free
parameters in the tensorial order-parameter field. For a given point
$P$, the plane $\Pi_P$ passing through $P$ and the $z$-axis is a
symmetry plane. Thus, the frame
$(\mathbf{e}^{(1)},\mathbf{e}^{(2)},\mathbf{e}^{(3)})$ in which $Q_{ij}$
is diagonal has two directions $\mathbf{e}^{(1)}$ and $\mathbf{e}^{(2)}$
in the plane $\Pi_P$ and the third, $\mathbf{e}^{(3)}$, normal to it.
This frame can be parametrized by the angle $\Theta(r,\theta)$ that
$\mathbf{e}^{(1)}$ makes with the
$z$-axis:
\begin{subequations}
\label{eq:e}
\begin{eqnarray}
\label{eq:e1}
\mathbf{e}^{(1)}&=&\sin\Theta\cos\phi\,\mathbf{\hat x}
+\sin\Theta\sin\phi\,\mathbf{\hat y}
+\cos\Theta\,\mathbf{\hat z},\\
\mathbf{e}^{(2)}&=&\cos\Theta\cos\phi\,\mathbf{\hat x}
+\cos\Theta\sin\phi\,\mathbf{\hat y}
-\sin\Theta\,\mathbf{\hat z},\\
\label{eq:e3}
\mathbf{e}^{(3)}&=&-\sin\phi\,\mathbf{\hat x}
+\cos\phi\,\mathbf{\hat y}.
\end{eqnarray}
\end{subequations}
Therefore, the order-parameter tensor $Q_{ij}$ can be written as 
\begin{equation}
\label{Q_deux}
Q_{ij}=\lambda_1\, e^{(1)}_i e^{(1)}_j
+\lambda_2\, e^{(2)}_i e^{(2)}_j
+\lambda_3\, e^{(3)}_i e^{(3)}_j,
\end{equation}
where its eigenvalues $\lambda_i$ satisfy
\begin{equation}
\label{eq:l3}
\lambda_3=-(\lambda_1+\lambda_2),
\end{equation}
since $Q_{ij}$ is traceless. The corresponding
$q$-components~(\ref{q-components}) are therefore
\begin{subequations}
\label{q-compo_pour2}
\begin{eqnarray}
q_0 &=& \alpha(r,\theta) \cos 2\phi,\\
q_1 &=& \beta(r,\theta) \sin\phi,\\
q_2 &=& \beta(r,\theta) \cos\phi,\\
q_3 &=& \frac{1}{2} \alpha(r,\theta) \sin 2\phi,\\
q_4 &=& \gamma(r,\theta),
\end{eqnarray}
\end{subequations}
with
\begin{subequations}
\label{eq:abc}
\begin{eqnarray}
\alpha(r,\theta)&=&\frac{\lambda_1}{S_0}\left(1+\sin^2\Theta\right)+
\frac{\lambda_2}{S_0}\left(1+\cos^2\Theta\right),\\
\beta(r,\theta)&=&\frac{\lambda_1-\lambda_2}{2S_0}
\sin2\Theta,\\
\gamma(r,\theta)&=&\frac{\lambda_1}{S_0}\cos^2\Theta+
\frac{\lambda_2}{S_0}\sin^2\Theta.
\end{eqnarray}
\end{subequations}
Hence the order-parameter field around two particles is fully described
by the three fields $\alpha(r,\theta)$, $\beta(r,\theta)$, and
$\gamma(r,\theta)$.

\subsubsection{Multipolar development}

For two particles, the most general expression of $Q_{ij}$ in terms of
multipoles must be a sum of \textit{two} multipolar expansions of the
kind~(\ref{expansion_multipolaire}), each one centered on each one of
the particles.  The reason is the following. Since a multipolar
expansion such as (\ref{expansion_multipolaire}) is singular at the
origin ($r=0$), the latter must be put inside one of the particles.
Moreover, using only one expansion, one would impose that the analytical
continuation of the texture is regular inside the other particle
(because the multipolar expansion is regular everywhere but at $r=0$),
which is not required.  Therefore, in the most general case, there must
be a second multipolar source inside the other particle. For two
identical particles, this is also obvious from symmetry considerations.

Taking into account the general form~(\ref{q-compo_pour2}) and the fact
that $(x,y)$ is a symmetry plane, the double multipolar expansion takes
the form (see also Appendix~\ref{sphharm}):
\begin{subequations}
\label{eq:expansion}
\begin{eqnarray}
\alpha(r,\theta) &=& \sum_{\ell=2}^\infty \sum_{p=1}^2 \alpha_\ell
u_\ell(r_p) {P_\ell}^2(\cos\theta_p), \\
\beta(r,\theta) &=& \sum_{\ell=1}^\infty \sum_{p=1}^2 (-1)^p \beta_\ell
u_\ell(r_p) {P_\ell}^1(\cos\theta_p),\\
\label{eq:gamma}
\gamma(r,\theta) &=& \sum_{\ell=0}^\infty \sum_{p=1}^2 \gamma_\ell
u_\ell(r_p) {P_\ell}^0(\cos\theta_p).
\end{eqnarray}
\end{subequations}
Thus, at this point, the whole order-parameter texture is uniquely
determined by the set of coefficients $\alpha_\ell$ (with
$\ell=2,3,\ldots$), $\beta_\ell$ (with $\ell=1,2,\ldots$), and
$\gamma_\ell$ (with $\ell=0,1,\ldots$).

\subsubsection{Boundary conditions}

Because of the presence of the $(x,y)$ symmetry plane, we need to impose
the boundary conditions~(\ref{eq:surface}) only on one of the particles,
let say particle~$1$. These boundary conditions on the $q$-components 
transform to boundary conditions on the fields $\alpha$, $\beta$
and~$\gamma$ [see Eqs.~(\ref{q-compo_pour2})], which in the coordinates
$(r_1,\theta_1)$ relative to particle~$1$, take the form
\begin{subequations}
\label{boundary2}
\begin{eqnarray}
\left.\frac{\partial\alpha}{\partial r_1}\right|_{r_1=R}&=&
w\left[\alpha(r_1\!=\!R,\theta_1)-\frac{1}{3} {P_2}^2(\cos\theta_1)
\right],\quad\\
\left.\frac{\partial\beta}{\partial r_1}\right|_{r_1=R}&=&
w\left[\beta(r_1\!=\!R,\theta_1)+\frac{1}{3} {P_2}^1(\cos\theta_1)
\right],\quad\\
\left.\frac{\partial\gamma}{\partial r_1}\right|_{r_1=R}&=&
w\left[\gamma(r_1\!=\!R,\theta_1)-\frac{2}{3} {P_2}^0(\cos\theta_1)
\right].\quad
\label{eq:bcg}
\end{eqnarray}
\end{subequations}
Note that these equations are decoupled; therefore the three fields
$\alpha$, $\beta$ and~$\gamma$ can be treated independently.

\subsubsection{Coordinate transformation}

To impose the boundary conditions~(\ref{boundary2}) to the multipolar
expansions~(\ref{eq:expansion}), we need to express the spherical
coordinates $r_2$ and~$\theta_2$ in terms of $r_1$ and~$\theta_1$. From
Fig.~\ref{2billes} we easily obtain
\begin{subequations}
\label{eq:trans}
\begin{eqnarray}
\label{eq:r2}
r_2 &=& \sqrt{{r_1}^2+d^2+2d\,r_1\,\cos\theta_1},\\
\label{eq:t2}
\cos\theta_2 &=&
-\frac{r_1\,\cos\theta_1+d}{\sqrt{{r_1}^2+d^2+2d\,r_1\,\cos\theta_1}}.
\end{eqnarray}
\end{subequations}

\subsection{Asymptotic solution}

The \textit{exact} two-particle order-parameter
texture~(\ref{eq:expansion}) satisfying the boundary
conditions~(\ref{boundary2}) cannot be determined analytically, owing to
the intricate relations~(\ref{eq:trans}) that link the local coordinate
systems centered on the two particles.  However, in the limit $d\gg1$,
i.e., for distances large with respect to the nematic coherence
length~$\xi$ (even if $d-2R\ll R$), we can obtain an \textit{asymptotic}
solution by expanding the unknown multipolar coefficients $\alpha_\ell$,
$\beta_\ell$, and $\gamma_\ell$ in series of~$e^{-d}/d^n$ up to a given
order~$n$.  In the following, we shall obtain the lowest order expansion,
$n=1$. It turns out that to this order the only non-zero terms in the
multipolar expansions~(\ref{eq:expansion}) are those up to $\ell=2$.

To begin with, we illustrate the strategy for computing the
$\gamma_\ell$ coefficients.  We start by writing the
expansion~(\ref{eq:gamma}) on particle~$1$
($r_1=R$) up to $\ell=2$:
\begin{widetext}
\begin{eqnarray}
\gamma(r_1=R,\theta_1) &=&
\gamma_0\left[
u_0(R) {P_0}^0(\cos\theta_1)+u_0(R_2) {P_0}^0(\cos\Theta_2)
\right]
+\gamma_1\left[
u_1(R) {P_1}^0(\cos\theta_1) +u_1(R_2) {P_1}^0(\cos\Theta_2)
\right]\nonumber\\
&+&\gamma_2\left[
u_2(R) {P_2}^0(\cos\theta_1) +u_2(R_2) {P_2}^0(\cos\Theta_2)
\right],
\label{eq:g1}
\end{eqnarray}
\end{widetext}
where $R_2=R_2(\theta_1)$ is the $r_2$ coordinate~(\ref{eq:r2})
evaluated on particle~$1$ ($r_1=R$), and, similarly,
$\cos\Theta_2=\cos\Theta_2(\theta_1)$ is the $\cos\theta_2$
coordinate~(\ref{eq:t2}) evaluated on particle~$1$. Next, we
asymptotically develop the functions $u_n(R_2) {P_n}^0(\cos\Theta_2)$ up
to terms of order~$e^{-d}/d$:
\begin{equation}
u_n(R_2) {P_n}^0(\cos\Theta_2) \simeq (-1)^n \frac{e^{-d}}{d}
e^{-R\cos\theta_1},
\end{equation}
and we redevelop this expansion on the complete basis of Legendre
functions of first kind ${P_n}^0(\cos\theta_1)$, using the orthogonality
relations~(\ref{eq:orto}).  Truncating the expansion at $\ell=2$, for
the sake of consistency, we arrive at 
\begin{widetext}
\begin{eqnarray}
u_n(R_2){P_n}^0(\cos\Theta_2) &\simeq& (-1)^n \frac{e^{-d}}{d}
\left\{\frac{\sinh R}{R} {P_0}^0(\cos\theta_1)
-3\left[\frac{\cosh R}{R}-\frac{\sinh R}{R^2} \right]
{P_1}^0(\cos\theta_1)\right.\nonumber\\
&+&5\left.\left[\frac{\sinh R}{R}-\frac{3\cosh R}{R^2}
+\frac{3\sinh R}{R^3}\right]{P_2}^0(\cos\theta_1)\right\}.
\label{eq:unpn}
\end{eqnarray}
Replacing Eq.~(\ref{eq:unpn}) into Eq.~(\ref{eq:g1}) yields the required
expansion of $\gamma$. To match the boundary conditions, let us
determine the corresponding expansion for the radial derivative
of~$\gamma$. To this aim, it is most efficient to take into account that
the radial functions~(\ref{ui}) obey the relations
\begin{equation}
\frac{d u_n(r)}{dr} = \frac{n u_n(r)}{r}-u_{n+1}(r).
\end{equation}
We then obtain
\begin{eqnarray}
\left.
\frac{\partial\gamma}{\partial r_1}\right|_{r_1=R} &=&
\gamma_0\left\{
-u_1(R) {P_0}^0(\cos\theta_1)+\frac{\partial}{\partial r_1}\left[u_0(r_2)
{P_0}^0(\cos\theta_2)\right]_{r_1=R}
\right\}\nonumber\\
&+&\gamma_1\left\{
\left[\frac{u_1(R)}{R}-u_2(R)\right]{P_1}^0(\cos\theta_1)
+\frac{\partial}{\partial r_1}\left[u_1(r_2)
{P_1}^0(\cos\theta_2)\right]_{r_1=R}
\right\}
\nonumber\\
&+&\gamma_2\left\{
\left[\frac{2\,u_2(R)}{R}-u_3(R)\right]{P_2}^0(\cos\theta_1)
+\frac{\partial}{\partial r_1}\left[u_2(r_2)
{P_2}^0(\cos\theta_2)\right]_{r_1=R}
\right\},
\label{eq:dg1}
\end{eqnarray}
in which, according to  Eq.~(\ref{eq:unpn}),
\begin{eqnarray}
\frac{\partial}{\partial r_1}\left[u_n(r_2) {P_n}^0(\cos\theta_2)
\right]_{r_1=R}&\simeq& (-1)^n \frac{e^{-d}}{d}
\left\{\left[\frac{\cosh R}{R}-\frac{\sinh R}{R^2} \right]
{P_0}^0(\cos\theta_1)\right.\nonumber\\
&-&3\left.\left[\frac{\sinh R}{R}-\frac{2\cosh R}{R^2}+\frac{2\sinh R}{R^3} \right]
{P_1}^0(\cos\theta_1)\right.\nonumber\\
&+&5\left.\left[\frac{\cosh R}{R}-\frac{4\sinh R}{R^2}
+\frac{9\cosh R}{R^3}-\frac{9\sinh R}{R^4}\right]{P_2}^0(\cos\theta_1)\right\}.
\label{eq:dunpn}
\end{eqnarray}
Finally, we insert the relations (\ref{eq:g1}), (\ref{eq:unpn}),
(\ref{eq:dg1}), and~(\ref{eq:dunpn}) in the boundary
condition~(\ref{eq:bcg}). Because of the orthogonality
relations~(\ref{eq:orto}), we get three linear coupled equations for the
unknown coefficients $\gamma_0$, $\gamma_1$, and~$\gamma_2$. Solving
them to leading order in $e^{-d}/d$ yields, after some lengthy algebra,
\begin{subequations}
\label{eq:gs}
\begin{eqnarray}
\gamma_0 &\simeq& \frac{e^{-d}}{d} \frac{e^R R^4 w\left\{
R\left(w+1\right)+1-e^{2R}\left[R\left(w-1\right)+1\right]
\right\}}{3\left[
R^4\left(w+1\right)^2+R^3\left(3w^2+8w+5\right)+R^2\left(
3w^2+15w+13\right)+ 6R\left(2w+3\right)+9\right]},\\
\gamma_1 &\simeq& \frac{e^{-d}}{d} \frac{e^R R^4 w\left\{
R^2\left(w+1\right)+R\left(w+2\right)+2
+e^{2R}\left[R^2(w-1)-R\left(w-2\right)-2\right]
\right\}}{R^5\!\left(w\!+\!1\right)^2+2R^4\!\left(2w^2\!+\!5w\!+\!3\right)
+R^3\!\left(6w^2\!+\!24w\!+\!19\right)+R^2\!\left(3 w^2\!+\! 30w\!+\!35\right)
+3R\!\left(5w\!+\!12\right)\!+\!18},\qquad\\
\gamma_2 &\simeq& \frac{2}{3}\mathcal{A}(R,w)e^{R}\left\{1+\frac{5}{2}
\frac{e^{-d}}{d}\left[1-\frac{e^{2R} \mathcal{A}(R,w)\left[
R^3\left(w-1\right)-R^2\left(3w-4\right)+3R\left(w-3\right)+9
\right]}{R^4w}
\right]
\right\}.\qquad
\end{eqnarray}
\end{subequations}
\end{widetext}

The calculation of the remaining coefficients $\alpha_\ell$
and~$\beta_\ell$ is much simpler since, to
leading order in $e^{-d}/d$, they do not depend on~$d$, because of the
asymptotic behavior
\begin{subequations}
\begin{eqnarray}
u_n(R_2) {P_n}^1(\cos\Theta_2) &\propto& \frac{e^{-d}}{d^2},\\
u_n(R_2) {P_n}^2(\cos\Theta_2) &\propto& \frac{e^{-d}}{d^3}.
\end{eqnarray}
\end{subequations}
These coefficients are therefore given by their expressions for an
isolated particle
\begin{subequations}
\begin{eqnarray}
\alpha_2 &\simeq& \frac{1}{3}\mathcal{A}(R,w)e^{R},\\
\beta_1 &\simeq& 0,\\
\beta_2 &\simeq& -\frac{1}{3}\mathcal{A}(R,w)e^{R}.
\end{eqnarray}
\end{subequations}

\subsubsection{Asymptotic interaction energy}
\label{sec:asympt}

To calculate the free-energy we use Eq.~(\ref{eq:fen}). By symmetry, it
is equal to twice the contribution corresponding to the surface of
particle~$1$. In order to calculate this contribution, we express
$\alpha(r_1=R,\theta_1)$, $\beta(r_1=R,\theta_1)$ and
$\gamma(r_1=R,\theta_1)$ as an expansion in Legendre functions: 
\begin{subequations}
\begin{eqnarray}
\alpha(r_1=R,\theta_1) &=& \sum_{n=2}^\infty A_n {P_n}^2(\cos\theta_1),\\
\beta(r_1=R,\theta_1) &=& \sum_{n=1}^\infty B_n {P_n}^1(\cos\theta_1),\\
\gamma(r_1=R,\theta_1) &=& \sum_{n=0}^\infty C_n {P_n}^0(\cos\theta_1).
\end{eqnarray}
\end{subequations}
Using Eq.~(\ref{eq:fen}) and the orthogonality of the Legendre
functions, we obtain 
\begin{equation}
F = \frac{4}{15}\pi R^2 w\left(10-12A_2+12B_2-3C_2\right).
\end{equation}
The coefficients $A_2$, $B_2$ and~$C_2$ are easily calculated from the
asymptotic expansions of $\alpha$, $\beta$, and~$\gamma$ given in the
preceding section.  We arrive at $F=2\,F_1+F_\mathrm{int}^\infty$, where
$F_1$ is the free-energy~(\ref{eq:f1}) of an isolated particle and
\begin{eqnarray}
F_\mathrm{int}^\infty &=& \frac{-8\pi R^8w^2 e^{-(d-2R)}}{3d
\left[
R^3\!\left(w\!+\!1\right)+R^2\!\left(3w\!+\!4\right)
+ 3R\!\left(w\!+\!3\right)\!+\!9\right]^2} \nonumber \\
 & = & -\frac{8\pi}{3}\, \mathcal{A}^{2}\, \frac{e^{-(d-2R)}}{d}
\label{5.40}
\end{eqnarray}
is the asymptotic interaction free-energy. It is always attractive and
has the form of a Yukawa potential~\cite{YukawaPotential}. In
Subsection~\ref{inter.num.ener}, using a numerical calculation, we shall
show that this asymptotic expression is quite good even up to
separations of the order of~$\xi$ or less.

\subsubsection{Analysis of the effective Yukawa potential}
\label{sec:yuk}

\begin{figure}[t]
\includegraphics[width=.6\columnwidth]{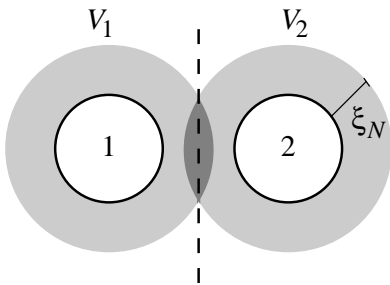}
\caption[]{The attractive paranematic interaction between the particles
comes from the overlapping of the two paranematic coronas of thickness
$\xi$.  In the geometric view, the interaction energy is equal to the
negative of the free-energy of the excess orientational order in the
dark shaded region [for an exact definition see Eq.~(\ref{5.50})]. The
half spaces $V_{1}$ and $V_{2}$ are defined by the midplane of particle
1 and 2.}
\label{fig.schem}
\end{figure}

There exists an instructive geometric view on the origin of the
attractive part of the two-particle interaction mediated by the
surface-induced paranematic order. It reproduces the form of the 
Yukawa potential of Eq.~(\ref{5.40}) under the assumption of 
large  particles ($R \gg 1$) and was presented by one of the 
authors in Ref.~\cite{Stark2002}. The basic idea is that the overlapping
of the paranematic coronas of the two particles reduces the volume
of the energetically unfavorable liquid crystal ordering, as 
illustrated in Fig.~\ref{fig.schem}, and therefore the total
free-energy. The interaction energy is the free-energy of the removed
orientational order. If we denote by $Q^{(n)}_{ij}(\bm{r})$ the
single-particle order-parameter field centered on particle $n$ and
switch to the representation by the components $q_{i}^{(n)}$
($i=0,\ldots,4$) [see Eq.~(\ref{q-solution_1particule})], the
interaction energy is calculated as
\begin{equation}
F_{\mathrm{int}} = -2 \sum_{i=0}^{4} c_{i} 
\left\{ \int_{V_{2}} \left[
(q_{i}^{(1)})^{2} + (\nabla q_{i}^{(1)})^{2}
\right] d^{3}r
\right\} \enspace,
\label{5.50}
\end{equation}
where $V_2$ denotes the half space of particle 2 
(see Fig.~\ref{fig.schem}). This definition leads to the 
Yukawa potential of Eq.~(\ref{5.40}) in the limit $R \gg 1$,
however with half the strength. Its advantage lies in its
semiquantitative agreement with Eq.~(\ref{5.40}) and its simplicity.
Using a type of Voronoi cell construction, it is extensible to
multibody interactions which are important in the study of particle
aggregation and the formation of ordered crystalline structures.

It was also demonstrated in Ref.~\cite{Stark2002} that the Yukawa potential
of Eq.~(\ref{5.40}) can exactly be derived by approximating the 
tensorial order-parameter through a linear superposition of the two 
single-particle solutions centered on particle 1 and 2
\begin{equation}
Q_{ij}(\bm{r}) = Q^{(1)}_{ij}(\bm{r}) + Q^{(2)}_{ij}(\bm{r}) \enspace.
\end{equation}
This approach is in full analogy to the treatment of the two-particle
potential in electrostatically stabilized colloids based on the 
Debye-H\"uckel approximation.

\subsubsection{Derjaguin approximation}
\label{sec:derj}

The force between two interacting spheres is often calculated in terms
of the interaction energy per unit surface~$\mathcal{E}(s)$ between two
parallel plates a distance~$s$ apart, using the so-called Derjaguin
approximation~\cite{Israelachvili92}. For two equal spheres of radii~$R$
a distance $s$ apart, the force in the Derjaguin approximation is simply
$\mathcal{F}_D(s) = \pi R \mathcal{E}(s)$~\cite{Israelachvili92}. Such
an approximation is valid if the interactions are additive and if the
radii of the particles are large with respect to both the range of the
interaction and the minimum separation distance~$s$ between the
particles. Although in our case the interaction energy is not
pairwise-additive, let us examine whether there exists a regime in which
the Derjaguin approximation holds.

\begin{figure}[t]
\includegraphics[width=\columnwidth]{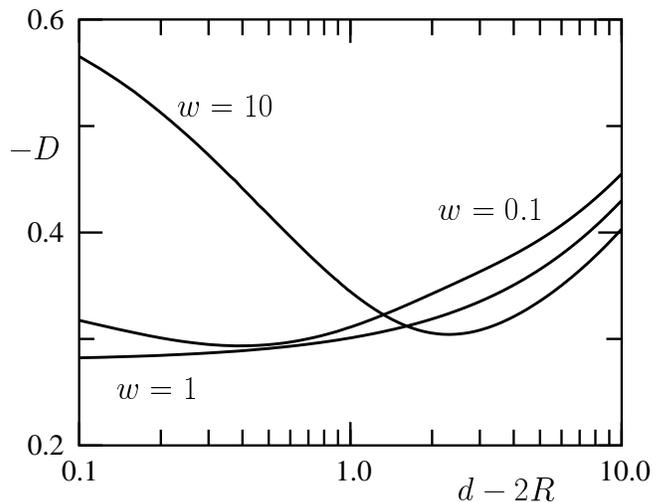}
\caption{Relative error $D$ of the force in the Derjaguin approximation
with respect to the exact numerical one, calculated according to the
procedure outlined in Section~\protect\ref{subsec.inter.num}.  The error
is plotted as a function of the distance to contact~$s=d-2R$ for a
particle of reduced radius $R=20$ and for various values of the reduced
anchoring~$w$. Contrary to what expected, the approximation does not
improve at short distances.}
\label{fig:derja}
\end{figure}

The interaction energy between two parallel plates, imposing homeotropic
boundary conditions, immersed in a nematic liquid crystal was calculated
in~\cite{borstnik97}. For weak induced paranematic order in the
isotropic phase, we can use our quadratic free-energy~(\ref{eq:qfe}).
The resulting order-parameter is everywhere uniaxial, with a nematic
director everywhere orthogonal to the surfaces. In our reduced units, a
straightforward calculation gives the exact interaction energy per unit
surface
\begin{equation}
\mathcal{E}(s) = -\frac{4 w^2}{3\left(1+w\right) \left[
e^s\left(1+w\right)+w-1\right]},
\end{equation}
that is everywhere attractive.
The Derjaguin approximation gives the interaction force
\begin{equation}
\label{eq:derja}
\mathcal{F}_D(s) =
-\frac{4 \pi R w^2}{3\left(1+w\right) \left[e^s\left(1+w\right)+w-1\right]}.
\end{equation}

To check the Derjaguin approximation, we need to compare the
\textit{forces} and not the interaction energy. Indeed, since the
Derjaguin approximation does not hold for separations large with respect
to the radii of the particle, we cannot fix the correct zero level for
the interaction energy. We calculate the exact force $\mathcal{F} =
-\partial F_\mathrm{int}/\partial d$ by numerically evaluating the
interaction energy~$F_\mathrm{int}$, as will be explained in
Sec.~\ref{subsec.inter.num}. In Fig.~\ref{fig:derja} we plot the
relative error~$D$, defined through~$\mathcal{F} =
\mathcal{F}_D\left(1+D\right)$. As it is seen, the Derjaguin
approximation is rather crude. Moreover, it does not improve as the
separation distance $s$ tends to zero, as is normally expected. This is
most probably due to the non pairwise-additive character of the force.
The error is actually lowest in the intermediate range $1\ll s \ll R$.
Indeed, in the limit $s\gg1$ the Derjaguin
approximation~(\ref{eq:derja}) becomes \begin{equation} \mathcal{F}_D(s)
\simeq  -\frac{4\pi R w^2 e^{-s} }{3\left(1+w\right)^2}, \end{equation}
which coincides, in the limit $R\gg 1$, with the exact asymptotic force
$-\partial F_\mathrm{int}^\infty/\partial d$ obtained from the
asymptotic interaction energy~(\ref{5.40}), with $s=d-2R$. Care should
therefore be taken in using the Derjaguin approximation for distances to
contact comparable with the nematic coherence
length~$\xi$~\cite{kocevar01}.

\begin{figure}
\includegraphics[width=\columnwidth]{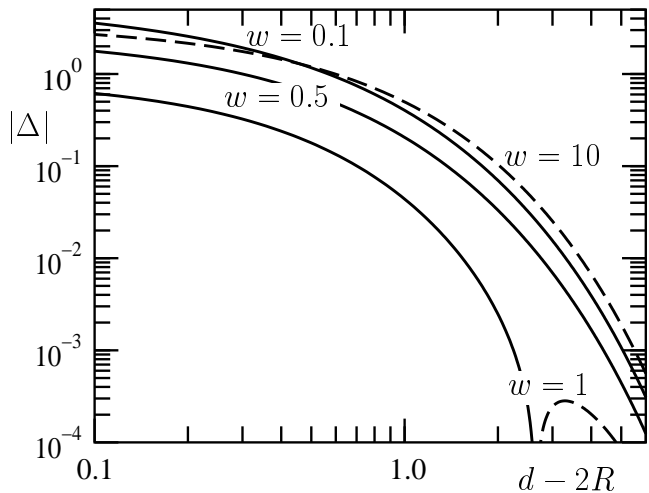}
\caption{Absolute value of the relative error $\Delta$ of the asymptotic
interaction energy as a function of the distance to contact~$d-2R$ for
$R=2$ and various values of the reduced anchoring~$w$. The solid lines
indicate that $\Delta>0$, the dashed lines that~$\Delta<0$.}
\label{fig:err2}
\end{figure}

\begin{figure}
\includegraphics[width=\columnwidth]{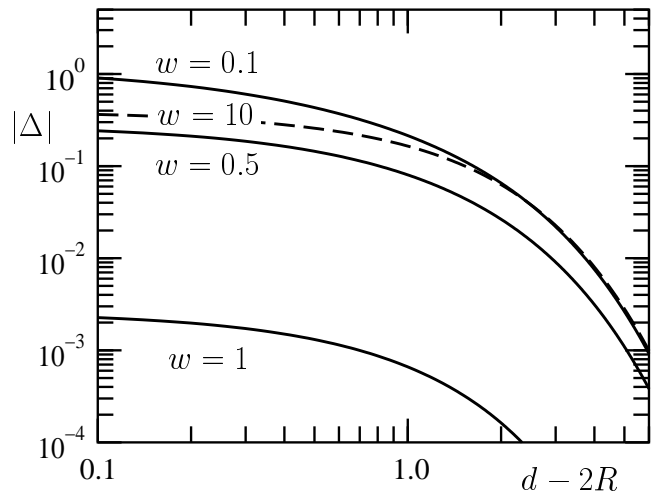}
\caption{As Fig.~\protect\ref{fig:err2} but for spheres of radius
$R=20$.}
\label{fig:err20}
\end{figure}

\subsection{Numerical results} \label{subsec.inter.num}

When the two colloidal particles are an arbitrary distance~$d$ apart, we
determine numerically the equilibrium order-parameter texture.  To this
aim, we truncate the expansions~(\ref{eq:expansion}) at some order
$\ell=\ell_\mathrm{max}$. Then, with the aid of the coordinate
transformation~(\ref{eq:trans}) and of the orthogonality
relations~(\ref{eq:orto}), we project numerically the boundary
conditions~(\ref{boundary2}) onto the Legendre functions of the first
kind ${P_\ell}^m(\theta_1)$, with $m=2$ for the function
$\alpha(r,\theta)$, $m=1$ for $\beta(r,\theta)$ and $m=0$ for
$\gamma(r,\theta)$, and $\ell=m,\ldots,\ell_\mathrm{max}$. This gives
three decoupled sets of linear equations that we solve numerically to
determine the unknown coefficients $\alpha_\ell$ (with
$\ell=2,\ldots,\ell_\mathrm{max}$), $\beta_\ell$ (with
$\ell=1,\ldots,\ell_\mathrm{max}$), and $\gamma_\ell$ (with
$\ell=0,\ldots,\ell_\mathrm{max}$). Finally, we vary the order
$\ell_\mathrm{max}$ of the expansions to obtain a given accuracy.
Typically, only a few multipoles ($\ell_\mathrm{max}\simeq 10\div 20$)
are needed to reach relative errors of the order of~$10^{-3}$ or less.

\subsubsection{Interaction energy} \label{inter.num.ener}

We determine the exact interaction energy of the colloidal particles by
inserting the numerically determined expansions~(\ref{eq:expansion}) in
the equilibrium free-energy~(\ref{eq:fen}). The behavior of the exact
interaction energy was studied in some detail in~\cite{galatola01}. Here
we concentrate on its comparison with the asymptotic
expression~(\ref{5.40}), to validate the latter. We set
\begin{equation}
F_\mathrm{int} = F_\mathrm{int}^\infty\left(1+\Delta\right),
\end{equation}
where~$\Delta$ measures the relative error from the asymptotic
solution. In Fig.~\ref{fig:err2} we show the behavior of $\Delta$
as a function of the distance to contact $d-2R$ for small particles
($R=2$) and various anchoring strength~$w$. As it is seen, the
asymptotic approximation is quite good up to distances to contact of the
order of~$\xi$. For low anchoring strength, $w<1$, the actual
interaction energy is larger than the asymptotic value. For high
anchoring strength, $w>1$, the actual interaction is smaller. The best
agreement is for anchoring strength of the order of~$w=1$, for which the
relative error increases up to $0.6$ at particle contact.

For relatively large particles ($R=20$), the relative error
$\Delta$ becomes even smaller, as shown in Fig.~\ref{fig:err20}.
It is remarkable that around $w=1$ the asymptotic approximation gives
extremely good results up to contact for larger spheres.

\subsubsection{Texture around two particles}
\label{sec:texture}

According to (\ref{eq:abc}) and~(\ref{eq:l3}), the eigenvalues
associated with the eigenvectors~(\ref{eq:e}) of the order-parameter
tensor are
\begin{widetext}
\begin{subequations}
\begin{eqnarray}
\label{eq:l1}
\frac{\lambda_1(r,\theta)}{S_0}
&=&\frac{\alpha(r,\theta)+\gamma(r,\theta)}{4}
+\sqrt{\left[\frac{\alpha(r,\theta)-3\gamma(r,\theta)}{4}\right]^2
+\beta^2(r,\theta)},\\
\label{eq:l2}
\frac{\lambda_2(r,\theta)}{S_0}
&=&\frac{\alpha(r,\theta)+\gamma(r,\theta)}{4}
-\sqrt{\left[\frac{\alpha(r,\theta)-3\gamma(r,\theta)}{4}\right]^2
+\beta^2(r,\theta)},\\
\frac{\lambda_3(r,\theta)}{S_0}&=&-\frac{\alpha(r,\theta)
+\gamma(r,\theta)}{2},
\end{eqnarray}
\end{subequations}
while the angle $\Theta(r,\theta)$ that the
eigenvector~$\mathbf{e}^{(1)}$ makes with respect to the $z$-axis is
\begin{equation}
\Theta(r,\theta) = \tan^{-1}\left[
\frac{\alpha(r,\theta)-3\gamma(r,\theta)+
\sqrt{\left[\alpha(r,\theta)-3\gamma(r,\theta)\right]^2
+16\beta^2(r,\theta)}}{4\beta(r,\theta)}
\right].
\end{equation}
\end{widetext}

Far from the particles, or for an isolated particle, the texture is
uniaxial, with $\lambda_2=\lambda_3$. The paranematic
director~$\mathbf{n}$ thus coincides with the
eigenvector~$\mathbf{e}^{(1)}$ [see Eq.~(\ref{eq:e1})] associated
with ~$\lambda_1$, which is the eigenvalue of largest absolute value.
Therefore, by continuity, we choose as paranematic director~$\mathbf{n}$
the direction~$\mathbf{e}^{(1)}$. In general, however, the tensor
order-parameter Q is biaxial and is characterized by the uniaxial
scalar order-parameter~$S$ in the direction of the paranematic director
and by the biaxiality~$B$ in the plane orthogonal to it. They are
defined by setting in the diagonal frame
$(\mathbf{e}^{(1)},\mathbf{e}^{(2)},\mathbf{e}^{(3)})$, in which
$Q=\mathrm{diag}(\lambda_1, \lambda_2, \lambda_3)$,
$Q=\mathrm{diag}(2/3\, S, -1/3\,S+B, -1/3\,S-B)$~\cite{deGennes_book}.
Therefore:
\begin{subequations}
\label{eq:sb}
\begin{eqnarray} S(r,\theta) &=&
\frac{3}{2}\lambda_1(r,\theta),\\ B(r,\theta) &=&
\frac{1}{2}\left[\lambda_1(r,\theta)+2\lambda_2(r,\theta)\right].
\end{eqnarray}
\end{subequations}

\begin{figure}
\includegraphics[width=\columnwidth]{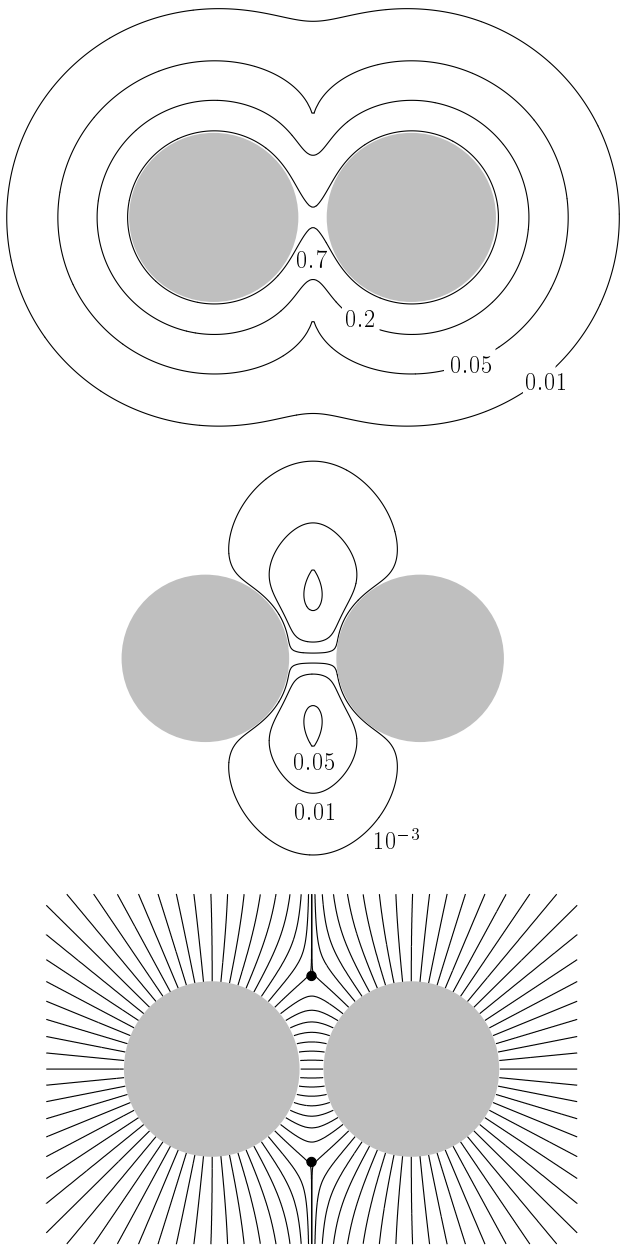}
\caption{Paranematic order-parameter between two spheres with reduced
radius $R=2$ and reduced anchoring $w=4$.  Top: contour lines of the
scalar order-parameter~$S$; middle: contour lines of the biaxiality
parameter $B$; bottom: field lines of the paranematic director
$\mathbf{n}$.}
\label{fig:texture2}
\end{figure}

\begin{figure}
\includegraphics[width=\columnwidth]{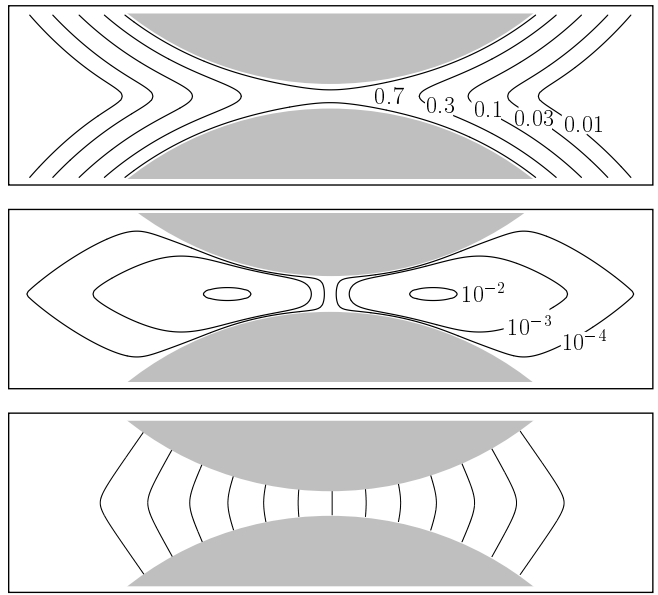}
\caption{Same as Fig.~\protect\ref{fig:texture2}, but for $R=20$.}
\label{fig:texture20}
\end{figure}

In Fig.~\ref{fig:texture2} we have plotted the contour lines of the
scalar order-parameters $S$ (top) and $B$ (middle), and the field-lines
of the paranematic director $\mathbf{n}$ (bottom) for small spheres,
i.e., $R=2$. In Fig.~\ref{fig:texture20} we have plotted the same
parameters for large spheres, i.e., $R=20$. When the two spheres are
close to each other, the scalar order-parameter makes a diffuse halo
around the two spheres for the small spheres, whereas it is
significantly nonzero only in the gap between the spheres for the
large ones. The biaxiality is maximum in the vicinity of a ring which
defines a defect, of topological charge $-1/2$, of the paranematic
director~$\mathbf{n}$. The defect location corresponds to the condition
$S=B$, i.e., $\lambda_1=\lambda_2$ [see Eqs.~(\ref{eq:sb})]: the two
orthogonal direction $\mathbf{e}^{(1)}$ and~$\mathbf{e}^{(2)}$ in the
symmetry plane $\Pi_P$ (the plane of Fig.~\ref{2billes} containing the
$z$-axis and the generic point~$P$) become equivalent.  Strictly
speaking, at the defect location the texture becomes again uniaxial,
with a paranematic director orthogonal to the $\Pi_P$-plane and a
discotic-like order. The main difference between the small and the
large spheres is that the defect is in the region where $S$ is large in
the former case, whereas it is located in a region where $S$ is almost
zero in the latter. Regarding the texture, one can see that the field
lines are rather smoothly bent in the case of small spheres, while
for large ones they start almost radially from the particles and
present a kink in the mid-plane, in the region where $S$ is small.

\subsubsection{Defect position}
\label{sec:defects}

The defect ring is located in the plane $z=0$ of Fig.~\ref{2billes}.
On this plane, $\beta=0$ for symmetry reasons. Then, the location of the
defect, which corresponds to the condition $\lambda_1=\lambda_2$,
according to Eqs.\ (\ref{eq:l1}) and~(\ref{eq:l2}) is such that
$\alpha(r,\theta=\pi/2)=3\gamma(r,\theta=\pi/2)$. Let us calculate
asymptotically, for $d\gg1$, the defect position.  At zeroth order, the
total $Q$ tensor is the superposition of the $Q$ tensors of the two
isolated particles. Then, one easily finds that the defect is located at
a radial distance $h=d/2$ from the axis joining the centers of the
particles.  This corresponds to a defect ring having a diameter equal to
the distance between the centers of the two spheres.

\begin{figure}
\includegraphics[width=\columnwidth]{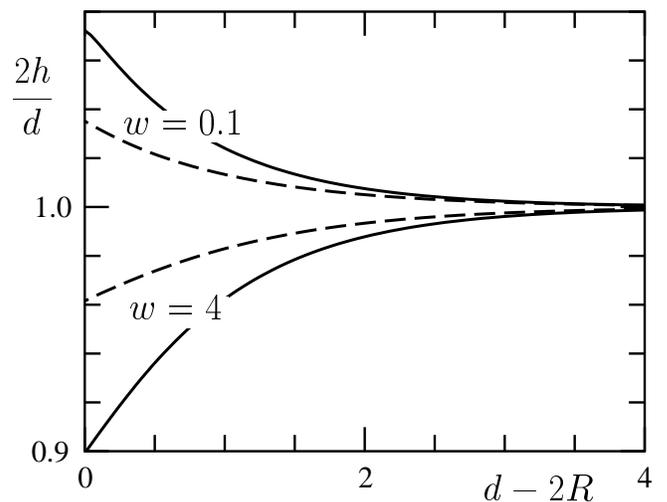}
\caption{Ratio between the defect ring diameter~$2h$ and the distance
between the spheres~$d$ as a function of the distance to contact~$d-2R$
of the spheres. Full lines: $R=2$; dashed lines: $R=5$. The two upper
curves are for a normalized anchoring strength $w=0.1$, the two lower
ones for $w=4$.}
\label{fig:posdef}
\end{figure}

To verify this lowest-order prediction, in Fig.~\ref{fig:posdef} we plot
the ratio between the ring diameter~$2h$ and the distance between the
two spheres as a function of the distance to contact~$d-2R$. As one
sees, the approximate estimation is rather good: for lower anchoring
strength, the defect diameter is actually slightly larger, while for
higher anchoring it is smaller.  The deviation from the lowest-order
solution $h=d/2$ decreases as the spheres move further apart or become
bigger. In this case, the defect ring lives in a region where the
order-parameter is small.

\section{Flocculation transition}
\label{sec:floc}

Section~\ref{inter.num.ener} (see also Ref.~\cite{fournier02}) and the
Yukawa potential of Eq.~(\ref{5.40}) show that for large particles ($R
\gg 1$), the two-particle interaction mediated by the nematic wetting
layer (in the following abbreviated by~$U_{LC}$) is dominated by its
strong attractive part. The consequences of this feature on the
stability of a dispersion of colloidal particles were investigated in
detail by Bor{\v s}tnik, Stark and {\v Z}umer in
Ref.~\cite{borstnik2000}.  The main idea was to use, e.g., the screened
electrostatic interaction of charged particles to stabilize a colloidal
suspension against the attractive van der Waals interaction well above
the nematic-isotropic phase transition. Then by lowering the temperature
towards the transition temperature $T_{c}$, which increases both the
strength and range of the liquid-crystal mediated interaction, a
flocculation transition can be induced, i.e., the particles start to
aggregate. Whereas the transition itself has not been observed so far,
the attractive two-particle potential has been demonstrated in excellent
experiments by Ko{\v c}evar and Mu{\v sevi{\v c} using an atomic force
microscope~\cite{Kocevar2001,Kocevar2002}.

In this article, we study again the flocculation transition for two
reasons. In Ref.~\cite{borstnik2000} the interaction potential~$U_{LC}$
was calculated by modeling the liquid-crystalline order with the help
of an ansatz function. A weak repulsive barrier in~$U_{LC}$ occurred
which seems to be an artifact in the construction of the ansatz
function. Secondly, by using the analytic form of the Yukawa potential
for~$U_{LC}$, the calculation of the total interaction energy becomes
much easier.  Therefore, in Section~\ref{subsec.floc.phase} we shall
discuss a flocculation diagram in terms of the relevant parameters of
the electrostatic interaction which covers a much larger range than in
Ref.~\cite{borstnik2000}.  We will, however, confirm the flocculation
diagram of Ref.~\cite{borstnik2000}.

\subsection{Two-particle interactions}

In the following we consider a colloidal dispersion of particles subject
to van der Waals ($U_{W}$), electrostatic ($U_{E}$), and liquid crystal
induced ($U_{LC}$) interactions, where $U_{W}$ and~$U_{E}$ are taken
according to the DLVO theory~\cite{Derjaguin1941,Verwey1948}. In a
first approach, we assume that the surface induced nematic order has no
effect on $U_{W}$ and~$U_{E}$, so that the total two-particle potential
can be written as $U_{\mathrm{tot}} = U_{LC} + U_{W} + U_{E}$.

To be concrete, we consider silica particles of radius $R=250 \,
\mathrm{nm}$ dispersed in the liquid crystal 4-$n$-octyl-4'-cyanobiphenyl
(8CB). The van der Waals interaction between identical particles is
always attractive. It reads~\cite{Russel1995}
\begin{equation}
\label{6.1}
U_{W} = -\frac{H}{6} \, \left[\frac{2R^{2}}{d^{2}-4R^{2}} + 
\frac{2R^{2}}{d^{2}} + \ln\left(\frac{d^{2}-4R^{2}}{d^{2}}\right)
\right],
\end{equation}
where $d$ is the center-of-mass separation of the particles and
$H=1.1\,k_{\mathrm{B}}T$ is the Hamaker constant for silica particles
suspended in the compound 8CB~\cite{borstnik2000}. Strictly speaking,
via retardation effects, the Hamaker constant also depends on the
separation~$d$~\cite{Russel1995}. However, in our problem this
dependence is not crucial~\cite{borstnik2000}. Note that in the
following \textit{we do not use reduced units}: all physical quantities
keep their dimensions.

We stabilize the colloidal dispersion by electrostatic repulsion. The
particles possess an electrical surface charge density~$q_{s}$, and
monovalent salt of concentration~$n_{p}$ is added to the solvent. This
results in the two-particle potential~\cite{Russel1995}
\begin{equation}
\label{6.2}
U_{E} = - \frac{8\pi^{2}}{\varepsilon_{2}} \, \frac{R q_{s}^{2}}{\kappa^{2}}
\, \ln \left(1-e^{-\kappa(d-2R)} \right).
\end{equation}
The range of the interaction is determined by the Debye
length~$\kappa^{-1} = [\varepsilon_{2} k_{\mathrm{B}}T/(8\pi
e^{2}n_{p})]^{1/2}$, where~$\varepsilon_{2}$ denotes the dielectric
constant of the solvent and~$e$ is the fundamental charge. Note that
Eq.~(\ref{6.2}) is derived in the Derjaguin approximation under the
assumption $d-2R,\kappa^{-1} \ll R$. For realistic values of
$\kappa^{-1}$ and~$q_{s}$, see Ref.\ \cite{borstnik2000}
and~\cite{Kocevar2002}.

To include the liquid-crystal mediated interaction, we use the Yukawa
potential of Eq.~(\ref{5.40}). The latter, restoring to all the physical
quantities their dimensions, is given explicitly by
\begin{equation}
U_{LC} = -\frac{8\pi}{3} \, L \xi^2 (S_{0}\mathcal{A})^{2} \, 
\frac{e^{-(d-2R)/\xi }}{d},
\label{6.3}
\end{equation}
where the amplitude $\mathcal{A}$ must be calculated according to
Eq.~(\ref{4.10}), with $R$ substituted by $R/\xi$ and with $w=W / [L
\alpha (T-T^{\ast})]^{1/2}$ the reduced surface coupling constant. We
recall also that the nematic coherence length is given by $\xi =
[L/(\alpha(T-T^{\ast}))]^{1/2}$. In Section~\ref{subsec.inter.num} it
was shown that for large particle radii the Yukawa potential agrees well
with the numerical results down to a distance to contact of the order
of~$\xi$.  Even for smaller distances it gives a good approximation to
the interaction energy. We use Eq.~(\ref{6.3}) to obtain an estimate for
the strength of the Yukawa potential in the case~$R \gg \xi$:
\begin{equation}
|U_{LC}(d=2R)| \simeq \frac{4\pi}{3} \, L R S_{0}^{2} \, 
\left(\frac{w}{w+1}\right)^{2}.
\label{6.4}
\end{equation}
The rational function in the reduced surface coupling constant~$w$ is
monotonously growing like~$w^{2}$ for small~$w$ and approaching one at
large~$w$.  Therefore, we find that the strength of the Yukawa potential
increases, like its range~$\xi$, when the temperature is lowered towards
the nematic-isotropic phase transition at $T_{c}$. For example, for the
material parameters chosen immediately below, we find that the strength
increases by a factor of six when cooling down from $T_{c} +
10\,\mathrm{K}$ to~$T_{c}$.  When discussing the effect of the
interaction~$U_{LC}$, we shall use the parameters of 8CB [$\alpha=0.12
\times 10^{7}\,\mathrm{erg/(cm^{3}K)}$, $L=1.8 \times 10^{-6} \,
\mathrm{dyn}$, $T_{c}-T^{\ast} = 1.3 \, \mathrm{K}$, and $T_{c} =
314.8\,\mathrm{K}$], which give a coherence length $\xi(T_{c}) =
10.74\,\mathrm{nm}$ at the phase transition. Furthermore, we choose $W =
1.25 \, \mathrm{erg/cm^{2}}$ and~$S_{0} = 0.45$, in accord with
Ref.~\cite{borstnik2000}.  Finally, all interactions in the next
subsection are referred to the thermal energy $k_{\mathrm{B}}T$ at room
temperature, and, as already mentioned, the particle radius is $250\
\mathrm{nm}$.

\begin{figure}
\includegraphics[width=.9\columnwidth]{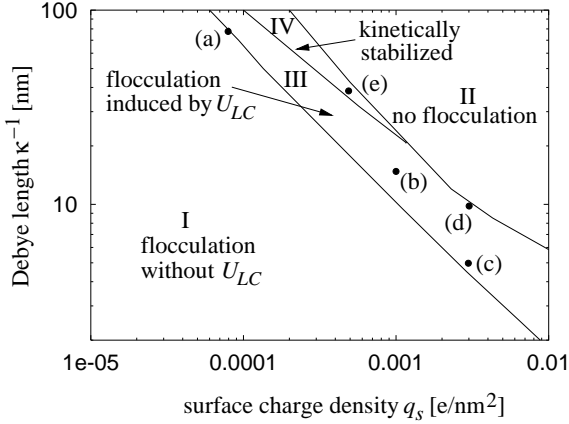}
\caption[]{Flocculation diagram in terms of the tunable parameters of 
the electrostatic interaction, i.e., the surface charge density 
$q_{s}$ and the Debye length $\kappa^{-1}$. The four different regions
I--IV characterize how a colloidal dispersion of charged particles
reacts on $U_{LC}$. The points (a)--(e) refer to the diagrams in
Figs.\ \ref{fig.inter2} and~\ref{fig.inter1}.}
\label{fig.phase}
\end{figure}

\begin{figure*}
\includegraphics[width=1.6\columnwidth]{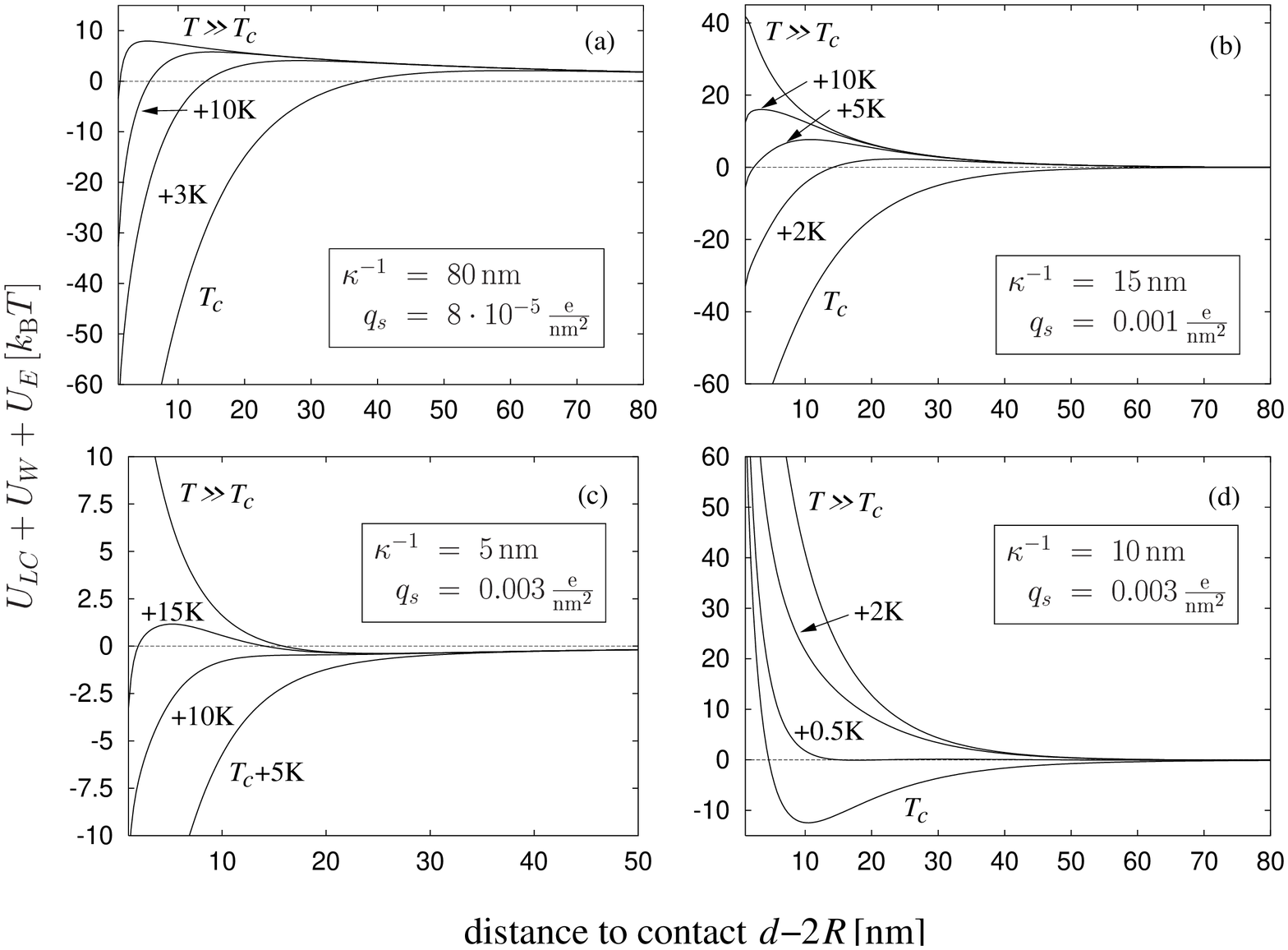}
\caption[]{Total two-particle potential in units of $k_{\mathrm{B}}T$
as a function of distance to contact $d-2R$. The potentials are shown 
for different parameters of the electrostatic interaction, as indicated 
in the inset, and at various temperatures relative to $T_{c}$. The labels
(a)--(d) refer to a location in the flocculation diagram of 
Fig.~\ref{fig.phase}.}
\label{fig.inter2}
\end{figure*}

\subsection{Flocculation diagram} \label{subsec.floc.phase}

When discussing the total interaction energy for varying parameters, we
will find different shapes of the two-particle potential which affect
the flocculation transition.  Particles start to aggregate when the
interaction potential exhibits a minimum~$U_{\mathrm{min}} <0$ at finite
distances.  If the potential minimum is shallow, i.e., just several
$k_{\mathrm{B}}T$ deep, a particle doublet will break up again, and a
phase equilibrium of an aggregated and dispersed state
occurs~\cite{Russel1995}. The higher potential energy in the dispersed
state is compensated by the larger entropy. On the other hand, when
$|U_{\mathrm{min}}|$ tends to~$10 \, k_{\mathrm{B}}T$ or even higher
values, strong attraction occurs that leads to a non-equilibrium phase
with all the particles aggregated.  This corresponds to the flocculation
transition that we aim to investigate. As a further feature of our
interaction potentials, we will also encounter potential barriers which
slow down the flocculation or even ``kinetically stabilize'' the
dispersion.

Obviously, the observation of a flocculation transition is a matter
of time scale on which the experiments take place. The theory of
aggregation kinetics determines the characteristic time~$\tau$ for the 
formation of a particle doublet as~\cite{Prieve1980,Russel1995}
\begin{equation}
\tau = \frac{1}{\phi}\frac{R^{2}}{6 D_{0}} \, I ,
\label{6.5}
\end{equation}
where the factor~$I$ is given by
\begin{equation}
I = 2R \int_{s_{\mathrm{min}}}^{\infty} \frac{D_{0}}{D(s)}
\frac{\exp(U_{\mathrm{tot}}/k_{\mathrm{B}}T)}{(2R+s)^{2}}\, ds ,
\label{6.6}
\end{equation}
which contains the ratio
\begin{equation}
\frac{D(s)}{D_{0}} = 
\left\{ \begin{array}{lll}
         2s/R & \mathrm{for} & s \ll R \\
           1   & \mathrm{for} & s \gg R
       \end{array}\right. .
\end{equation}
In Eq.~(\ref{6.5}), $\phi$ is the volume fraction of the particles in
the solvent, and $D_{0} = k_{B}T / (6\pi \eta R)$ is the single particle
diffusion coefficient depending on the solvent viscosity~$\eta$. For
particles that stick together when they meet but otherwise do not
interact ($I=1$), $\tau$ is given by the time an independent particle
needs to diffuse a distance~$R$ times the inverse volume
fraction~$\phi$. Clearly, in a very dilute dispersion ($\phi$ small),
particles hardly come close to each other and doublet formation is very
rare. The factor $I$, where $s=d-2R$ denotes the distance to contact,
incorporates influences from the two-particle interaction; in the
presence of a potential barrier, $\tau$ increases considerably even for
moderate~$\phi$. The ratio $D_{0}/D(s)$ takes into account corrections
of the diffusion constant due to hydrodynamic interactions.  Note that
$s_{\mathrm{min}}$ generally means $s=0$, i.e., the particles
stick together at direct contact. If, however, for very small particle
separations a strong repulsion sets in, $s_{\mathrm{min}}$ is taken as
the location of the potential minimum which is responsible for the
formation of the particle doublet.

In the following, we assume a volume fraction $\phi = 0.1$ so that for a
typical shear viscosity of $0.4\,\mathrm{P}$ we arrive at $\tau / I =
5\,\mathrm{s}$. We say that a dispersion is kinetically stabilized by a
potential barrier if $\tau > 1\,\mathrm{h}$, which corresponds
to~$I=720$.  With the help of the algebraic program Maple, we have
numerically calculated the factor~$I$ for our potential
$U_{\mathrm{tot}}$ to check when such a kinetic stabilization in the
flocculation diagram, to be discussed below, sets in. In this way, we
have determined the transition lines between regions I/III and III/IV in
Fig.~\ref{fig.phase}.  To gain further insight into the factor~$I$, we
perform a saddle-point approximation. We replace $U_{\mathrm{tot}}$ by
its harmonic approximation around the maximum $U_{\mathrm{max}}$ of the
potential barrier. Then we can evaluate the integral in Eq.~(\ref{6.6})
when we approximate $s$ by $s_{\mathrm{max}}$ otherwise and choose
$s_{\mathrm{min}} = 0$.  From the magnitude of the curvature $c_{0}= -
\partial^{2}U_{\mathrm{tot}}/\partial s^{2}|_{s_{\mathrm{max}}}$, we
deduce a characteristic length $\xi = (2k_{\mathrm{B}}T / c_{0})^{1/2}$
and make the approximation of replacing $s_{\mathrm{max}}$ by $\xi$
which finally gives the estimate $I =
\exp(U_{\mathrm{max}}/k_{\mathrm{B}}T) /2$. From our criterion for
kinetic stabilization, $I=720$, we find $U_{\mathrm{max}} = 7.3\,
k_{\mathrm{B}}T$. Surprisingly, in most cases the exact calculation of
the integral in Eq.~(\ref{6.6}) gave potential barriers with
$U_{\mathrm{max}}$ between 7 and $8\,k_{\mathrm{B}}T$. This demonstrates
that the Boltzmann factor is the determining quantity in~$I$ and
therefore in~$\tau$.

We will also encounter two-particle interactions where the particle
doublet settles into a potential minimum without traversing a potential
barrier. Here the question arises how stable the doublet is.
Unfortunately, the familiar Kramers rate~\cite{Kramers1940} is not
applicable, since it involves the curvature of the potential maximum over
which a particle escapes. In our case this potential maximum is located
at infinity ($U_{\mathrm{tot}} = 0$). It has zero curvature, which gives
an infinite escape time. The theory for our problem formulates the
escape time as a double integral over the particle
separation~\cite{Lifson1968}, that cannot be calculated
straightforwardly. However, we can give an upper bound for the escape
time~\cite{Lifson1968}, defined as the time $t_{\mathrm{diff}}$ that a
particle needs to diffuse a distance~$R$ when leaving a potential well
of depth~$U_{0}$
\begin{equation}
t_{\mathrm{diff}} = \frac{R^{2}}{6 D_{0}\exp(-U_{0}/k_{\mathrm{B}}T)} .
\end{equation}
Again, we consider a particle doublet as stable when $t_{\mathrm{diff}}$
exceeds $1\,\mathrm{h}$, which for the parameters already introduced
results in a potential depth of $U_{0}=9\,k_{\mathrm{B}}T$.  Compared to
the free diffusion time $R^{2}/(6D_{0}) = 0.5\,\mathrm{s}$, it enhances
$t_{\mathrm{diff}}$ by a factor of 7200.  We used this criterion to
establish the boundary line between region II and III in
Fig.~\ref{fig.phase}.

In the following, we discuss in detail the flocculation diagram of
Fig.~\ref{fig.phase} (calculated for a particle radius $R$ of
$250\,\mathrm{nm}$). It is plotted in terms of the tunable parameters of
the electrostatic interaction, i.e., the surface charge density~$q_{s}$
and the Debye length~$\kappa^{-1}$.  We distinguish four regions: in
region I the particles are completely aggregated without $U_{LC}$ since
the electrostatic interaction is not sufficient to stabilize the
dispersion against the van der Waals force.  For large surface charge
density and interaction range, as in region II, the dispersion is
completely stabilized even if $U_{LC}$ is turned on.  A potential
minimum at finite distances does not occur, or it is not deep enough,
according to our discussion above, to trigger flocculation.  Both
regions are separated by areas III and IV.  In region III flocculation is
induced by lowering the temperature towards~$T_{c}$. Different types of
interaction potentials occur which are illustrated in
Fig.~\ref{fig.inter2} for various locations in the flocculation diagram
of Fig.~\ref{fig.phase} labeled by (a) to (d).  In diagram~(a), the
graph for $T \gg T_{c}$ exhibits the usual ``primary minimum'' at short
distances due to the van der Waals interaction which is followed by a
``primary potential barrier'' for increasing separations well known in
colloids science. The electrostatic repulsion is just strong enough to
stabilize the colloidal suspension kinetically. Lowering temperature
reduces the barrier and induces flocculation.  Moving down the diagonal
from left to right in the flocculation diagram, we arrive at
location~(b). In the corresponding Fig.~\ref{fig.inter2}(b), the primary
minimum at $T \gg T_{c}$ is no longer visible due to the increase in the
surface charge density. To induce flocculation, one has to cool the
dispersion closer to~$T_{c}$. Moving further down the diagonal, a
shallow potential minimum at around $26\,\mathrm{nm}$ appears at~$T \gg
T_{c}$, as illustrated in Fig.~\ref{fig.inter2}(c), which leads to a
slight phase coexistence. However, a clear non-reversible aggregation of
the particles will be induced when the temperature is decreased.  A new
feature occurs in diagram~\ref{fig.inter2}(d), which is located close to
the transition line to the stabilized dispersion. Whereas in the
previous cases the flocculation transition sets in gradually with
decreasing temperature, here a sudden transition is observable within a
few tenths of Kelvin. Crossing the transition line from location (d),
the minimum at $T=T_{c}$ becomes more shallow and ultimately vanishes.
Region IV in the flocculation diagram of Fig.~\ref{fig.phase} identifies
a  kinetically stabilized dispersion.  As illustrated by
Fig.~\ref{fig.inter1}, at $T=T_{c}$ a minimum in the total interaction
energy is separated from large distances by a potential barrier which
prevents the formation of stable particle doublets.  The interaction
becomes totally repulsive when moving into region II.

\begin{figure}
\includegraphics[width=.9\columnwidth]{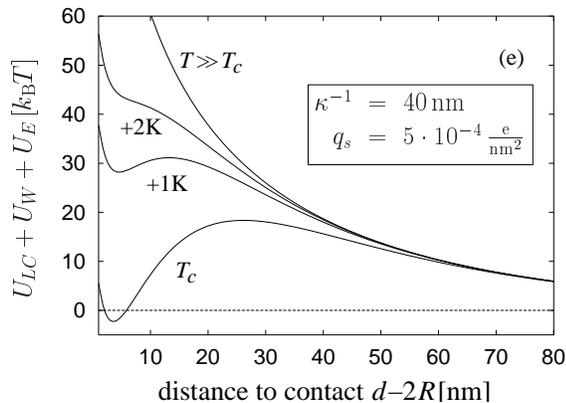}
\caption[]{Total two-particle potential in units of $k_{\mathrm{B}}T$
as a function of distance to contact $d-2R$. The potential is shown
at various temperatures relative to $T_{c}$. The parameters of the
electrostatic interaction is shown in the inset. The label~(e) refers
to a location in the flocculation diagram of Fig.~\ref{fig.phase}.}
\label{fig.inter1}
\end{figure}

\section{Conclusions}

In this paper, following previous
works~\cite{borstnik99,galatola01,Stark2002,borstnik2000,fournier02}, we
have reconsidered the interactions between spherical colloids wetted by
a corona of paranematic order in the isotropic phase of a nematogenic
material. We have calculated the contribution to the total interaction
that is mediated by the paranematic order, assuming a radial anchoring
of the director. Our results, based on a quadratic approximation of the
free-energy density, are exact in the limit of weak induced paranematic
order if the temperature is not too close to the nematic--isotropic
phase transition.  We have obtained analytical results in the asymptotic
regime where the distance between the colloids is large with respect to
the coherence length~$\xi$ of the nematic order, and numerical results
for any separation.  For large separations, the interaction follows the
Yukawa-like potential given by Eq.~(\ref{5.40}), which can be explained
on simple grounds. Our numerical results show that, for colloids large
compared to~$\xi$, this simple potential is actually a very good
approximation even up to separations of the order of~$\xi$. We obtain
attractive interactions for colloids much larger than~$\xi$ and the
possibility of short-range repulsions for colloids of size comparable
to~$\xi$. We have discussed the paranematic texture between two
colloidal particles, showing that a defect of the paranematic director,
in the form of a ring of topological charge $-1/2$ appears in the
midplane of the two particles. The diameter of the ring is approximately
equal to the distance between the centers of the two particles, which is
easily explained by supposing that at lowest order the total paranematic
tensor is the superposition of the paranematic tensors for the two
isolated particles. The ring is surrounded by a sheath of biaxial order.

Finally, using the Yukawa-like expression~(\ref{5.40}) for the paranematic
interaction, we have examined the stability of a colloidal suspension by
considering the interplay of the paranematic interaction with the
standard DVLO interactions, i.e., van der Waals attractions and
double-layer electrical repulsions.  We have found that the stability of
a colloidal suspension can be significantly affected by the
paranematic interaction: the latter can either trigger flocculation or
kinetically stabilize the suspension depending on the vicinity of the
isotropic--nematic transition of the nematogenic solvent and on the DLVO
parameters. These properties could be tested experimentally.

\appendix

\section{Definition of the spherical harmonics}\label{sphharm}
Among the various possible definitions of the spherical harmonics, we
use the form
\begin{equation}
Y_{\ell m}(\theta,\phi)=\sqrt{\frac{2\ell+1}{4\pi}
\frac{(\ell-m)!}{(\ell+m)!}}\,{P_{\ell}}^m(\cos\theta)\,e^{im\phi},
\end{equation}
where the ${P_{\ell}}^m(\cos\theta)$ are associated Legendre functions
of the first kind:
\begin{equation}
{P_{\ell}}^m(t) = \frac{(-1)^{\ell+m}}{2^\ell\,\ell!}\left(1-t^2\right)^{m/2}
\frac{d^{\ell+m}}{dt^{\ell+m}}\left(1-t^2\right)^\ell.
\end{equation}
The Legendre functions obey the orthogonality relations
\begin{equation}
\label{eq:orto}
\int_{-1}^1 {P_\ell}^m(t) {P_{\ell'}}^m(t)\, dt =
\frac{(\ell+m)!}{(\ell+1/2)(\ell-m)!}\delta_{\ell\ell'},
\end{equation}
where $\delta_{\ell\ell'}$ is the Kronecker delta. These orthogonality
relations imply the orthonormality of the spherical harmonics
\begin{equation}
\int Y_{\ell m}(\theta,\phi) Y^*_{\ell' m'}(\theta,\phi)\sin\theta
\, d\theta d\phi = \delta_{\ell \ell'} \delta_{m m'}.
\end{equation}

\bibliography{}

\end{document}